\documentclass[11pt]{article}\usepackage[]{graphicx}\usepackage[]{color}
\makeatletter
\def\maxwidth{ %
  \ifdim\Gin@nat@width>\linewidth
    \linewidth
  \else
    \Gin@nat@width
  \fi
}
\makeatother

\definecolor{fgcolor}{rgb}{0.345, 0.345, 0.345}

\usepackage{framed}
\makeatletter
 {\par\unskip\endMakeFramed%
 \at@end@of@kframe}
\makeatother

\definecolor{shadecolor}{rgb}{.97, .97, .97}
\definecolor{messagecolor}{rgb}{0, 0, 0}
\definecolor{warningcolor}{rgb}{1, 0, 1}
\definecolor{errorcolor}{rgb}{1, 0, 0}
\newenvironment{knitrout}{}{} % an empty environment to be redefined in TeX

\usepackage{alltt}
\usepackage{geometry} 

\usepackage{graphicx}
\usepackage{amssymb, amsmath}
\usepackage{epstopdf}
\usepackage{setspace}
\usepackage{parskip} % newlines before paragraphs
\raggedright 
\title{Infrastructure and methods for real-time predictions of the 2014 dengue fever season in Thailand}
\author{Nicholas G. Reich, Stephen A. Lauer, Krzysztof Sakrejda, \\ Sopon Iamsirithaworn, Soawapak Hinjoy, Paphanij Suangtho, Suthanun Suthachana, \\ Hannah E. Clapham, Henrik Salje, Derek A. T. Cummings, Justin Lessler }
\IfFileExists{upquote.sty}{\usepackage{upquote}}{}
\begin{document}
\maketitle

%\tableofcontents

\abstract{Epidemics of communicable diseases place a huge burden on public health infrastructures across the world. Producing accurate and actionable forecasts of infectious disease incidence at short and long time scales will improve public health response to outbreaks. However, scientists and public health officials face many obstacles in trying to create accurate and actionable real-time forecasts of infectious disease incidence. Dengue is a mosquito-borne virus that annually infects over 400 million people worldwide. We developed a real-time forecasting model for dengue hemorrhagic fever in the 77 provinces of Thailand.  We created an operational and computational infrastructure that generated multi-step predictions of dengue incidence in Thai provinces every two weeks throughout 2014. These predictions show mixed performance across provinces, out-performing na\"ive seasonal models in over half of provinces at a 1.5 month horizon. Additionally, to assess the degree to which delays in case reporting make long-range prediction a challenging task, we compared the performance of our real-time predictions with predictions made with fully reported data. This paper provides valuable lessons for the implementation of real-time predictions in the context of public health decision making.}

\clearpage

% \section*{Table and Figure summary}
% \begin{itemize}
%     %\item Table: Models fitted
%     \item DONE Table: list of delivery dates and analysis dates
%     \item DONE Figure: All data figure (maybe show aggregated country time-series to give context of 2014 being a low year?)
%     \item DONE Figure: Aggregated country predictions from several models at step 1
%     \item DONE Figure: Aggregated country predictions from several models at step 2
%     \item DONE Figure: 10-step predictions for 9 selected provinces
%     \item DONE Figure: map of outbreak probabilities
%     \item DONE Figure: relative MAE vs. seasonal
%     %\item Figure: For best prediction model, relative MAE vs. AR1
%     %\item Table: Country level metrics by model: pre-high-season predictions vs. observed number of high-season cases (country level), 
%     \item Table: province-level metrics.
% \end{itemize}

\section{Introduction}

%%% Overview
Producing accurate and actionable forecasts of infectious disease incidence at short and long time scales will improve public health response to outbreaks. %In recent years, there has been an increased interest in real-time predictions of infectious disease outbreaks. This is evidenced by several recent ``data competitions''. The US Centers for Disease Control and Prevention ran an influenza prediction challenge in [[2012]]. DARPA ran a competition to predict the spread of Chikungunya in the Americas during 2014. [[Run Web f Knowledge search to show increase?]] 
Real-time forecasts of infectious disease outbreaks can facilitate targeted intervention and prevention strategies, such as increased healthcare staffing or vector control measures. However, we currently have a limited understanding of the best ways to integrate forecasts into real-time public health decision-making. 

%%% Dengue background
Dengue is a mosquito-borne infectious disease that places an immense public health and economic burden upon countries around the world, especially in tropical areas. A severe form of the disease, dengue hemorrhagic fever (DHF), may lead to debilitating pain, organ shock, and even death \cite{Gubler:1998ws}. Currently over 2.5 billion individuals worldwide are at risk of infection with dengue, a mosquito-borne RNA virus \cite{WorldHealhOrganization:ww}. Global incidence of dengue has increased significantly over the past few decades, with estimated annual global incidence of about 400 million infections each year \cite{Bhatt:2013jb}. 

Dengue is endemic in Thailand, which has 77 provinces including one large municipality (Bangkok). National annual incidence rates of reported dengue in Thailand range between 30 cases per 100,000 population and 224 cases per 100,000 population \cite{Limkittikul:2014dh}. Some estimates suggest that between 50-80\% of cases may be inapparent and hence are difficult to detect clinically and often go unreported \cite{Endy:2011jg,Endy:2002vg,Endy:2010iu}. Annual outbreaks show dynamic temporal and spatial patterns, with great year-to-year and across-province variation, making public health planning and resource allocation an ongoing challenge \cite{Cazelles:2005es,Cummings:2004fy}.

%% Surveillance overview
With the maturation of disease surveillance and reporting systems in recent years, real-time disease forecasting has become a realistic goal in some settings.
Recognizing the importance of this emerging field, several governmental agencies have established disease prediction contests in recent years, with the goal of having contestants produce accurate forecasts: e.g. a 2013 CDC influenza prediction challenge \cite{CentersforDiseaseControlandPrevention:2013tf}, a 2014 DARPA chikungunya prediction challenge \cite{DefenseAdvancedResearchProjectsAgency:2014uy}, and a 2015 NOAA and White House dengue prediction challenge \cite{PandemicPredictionandForecastingScienceandTechnologyInteragencyWorkingGroup:uw}. 
However, researchers and practitioners are still working to understand and establish a set of best practices for implementing real-time prediction algorithms in practice. 

Creating predictions in real-time poses operational, computational, and statistical challenges. Operationally, raw data must be made available in a standard format for processing into analysis datasets. Historical data is also needed to allow for training of the prediction model(s). To enable transparent evaluations, predictions should be formally registered and archived in a publicly available database. Computational infrastructure is needed to transform and/or merge raw data into the analysis dataset and to run the models themselves. Analytical challenges include appropriate model training, selection, and validation, considering adjustments for delayed or incomplete case reporting. Depending on the methods used, additional statistical work may be necessary to accurately report uncertainty in the reported predictions. Below, we describe our approaches to dealing with these challenges.

%%% Our achievements
In this manuscript we present the results from the first year of forecasting DHF across the 77 provinces in Thailand. In 2014 our research team, a collaboration of the Ministry of Public Health of Thailand and researchers from multiple academic institutions, implemented a system for generating real-time forecasts of DHF based on current disease surveillance reports from Thailand. 
This paper illustrates several key components of a successful real time prediction framework, including: 
\begin{itemize}
\item a reliable pipeline for data transfer, cleaning, and analysis, with a data storage architecture that can recreate datasets that were available at a particular time (Section 2),
\item a statistical model of disease transmission used to generate real-time predictions of infectious disease incidence (Section 3),
\item an appropriate and rigorous model validation framework, including aggregating evaluations across location, calendar time, and prediction horizon (Section 4), and
\item an assessment of the impact of case reporting delays on the accuracy of predictions (Sections 3.3 and 4.2).
\end{itemize}
Valuable efforts have been made to create, validate, and operationalize real-time influenza predictions for the US \cite{Shaman:2013dr}, although these efforts have not faced the same challenges of systematic under-reported data. The operational infrastructure that we present in this manuscript provides valuable lessons for other collaborative prediction efforts between public health agencies and academic partners.

\section{A real-time prediction pipeline: turning data into forecasts}

\subsection{Data overview}
The data presented here come from the national surveillance system run by the Ministry of Public Health in Thailand. Monthly dengue hemorrhagic fever (DHF) case counts for each province are available from January 1968 through December 2005. Individual case reports (hereafter referred to as ``line-list'' data) were available for dengue fever (DF), DHF, and dengue shock syndrome from January 1, 1999 through December 31, 2014. The line-list data contains information on each case, including date of symptom onset, home address-code of the case (similar to a U.S. zip code), disease diagnosis code, and demographic information (sex, marital status, age, etc.). In years where we had overlapping sources for case data, the line-list data were used. A summary of province-level characteristics for all provinces in Thailand is provided in the appendix. Since 1968, several provinces have split into multiple provinces. Details on how we accommodate these province separations are available in the appendix. In one instance, the counts associated with a province (Bueng Kan) that split from another (Nong Khai) in 2011 have continued to be counted with the original province since we do not yet have enough data to predict for the new province.

Theoretical work demonstrates that by choosing the generation time as the discrete time interval for case reporting, the case reports may more easily be used to model the reproductive rate of the disease \cite{Nishiura:2010jh}. The generation time for dengue is two weeks, hence we aggregated the line-list data into biweekly intervals and interpolated the monthly counts into biweekly counts. (We used a definition of biweeks that followed a standardized definition based on calendar dates. See Supplemental Table 1.) Interpolation was performed by fitting a monotonically increasing smooth spline to the cumulative case counts in each province, and then taking the differences between the estimated cumulative counts at each interval as the number of incident cases in a given interval.

We chose to use only DHF cases because: (1) DHF is the only disease reported consistently across the 47 years of data collection, (2) DHF is less likely than DF to be misdiagnosed or to be differentially detected over time, and (3) from a public health perspective, DHF is a more relevant outcome, as it is a life-threatening condition and requires medical attention.

The research aspects of this study were approved by the Johns Hopkins Bloomberg School of Public Health and University of Massachusetts Amherst institutional review boards.

\subsection{Real-time data management}
We established a secure data transfer process to transmit data from the Thai disease surveillance system to U.S. researchers. Throughout the 2014 calendar year, Thai public health officials transmitted data approximately every two weeks to a secure server based in Baltimore, Maryland (Supplemental Table 2). These data were then loaded into a PostgreSQL database containing all of the data, including monthly case counts and a table with all line-list data received to date. The final report containing a cleaned and finalized record of all cases for the 2014 season was delivered in April 2015.  As of that time, this database held records of 2,586,928 unique cases of dengue in Thailand for the years 1968 through 2014, including records of 
1,930,858 
DHF cases (Figure \ref{fig:all-cases}).

% for all case counts
% sum(dhf_data$case_count, na.rm=TRUE)

\begin{knitrout}
\definecolor{shadecolor}{rgb}{0.969, 0.969, 0.969}\color{fgcolor}\begin{figure}
\includegraphics[width=\maxwidth]{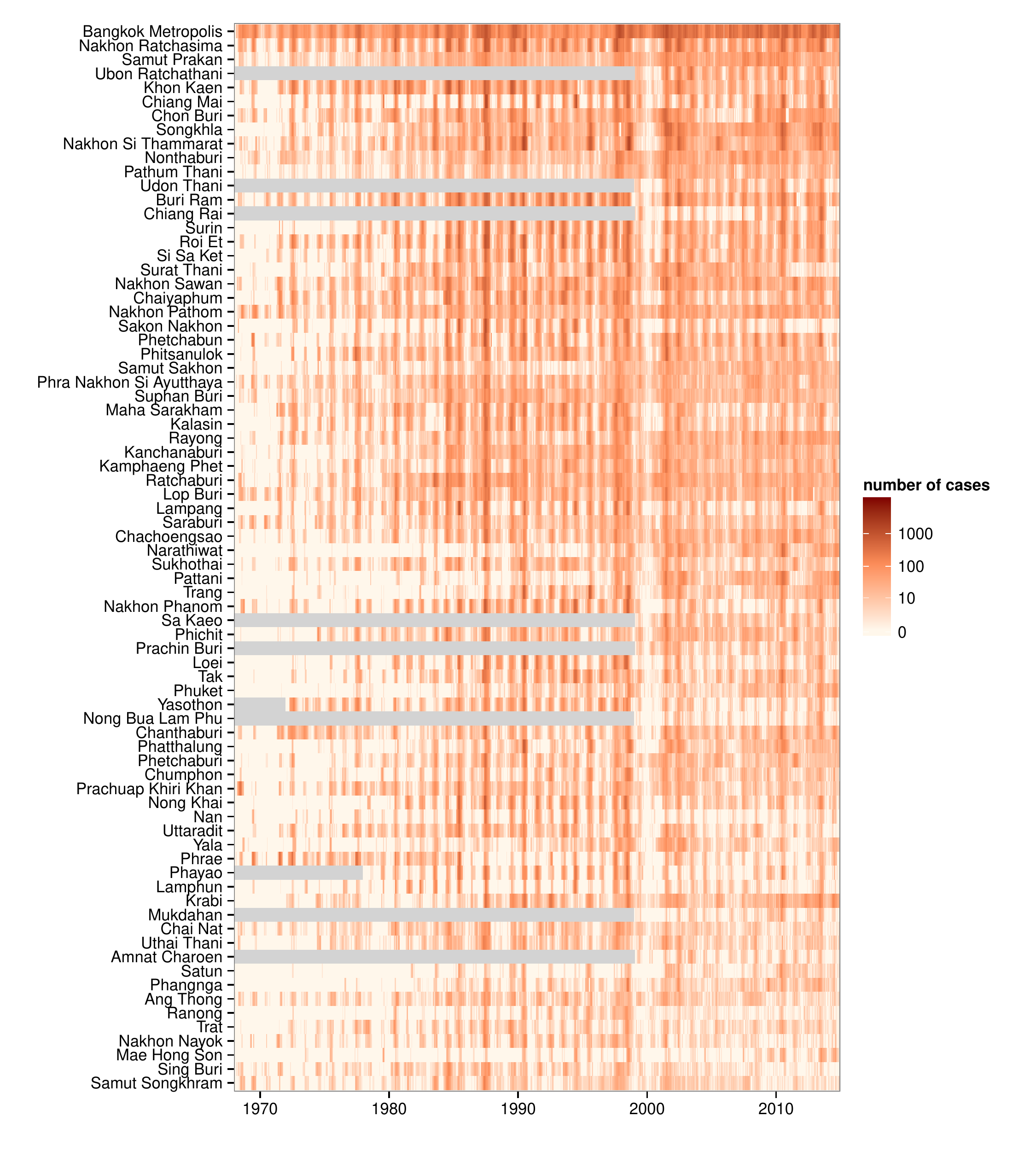} \caption[Raw dengue hemorrhagic fever case counts for 77 provinces of Thailand across 47 years (1968 - 2014)]{Raw dengue hemorrhagic fever case counts for 77 provinces of Thailand across 47 years (1968 - 2014). Provinces are ordered by by population (larger populations on the top). Gray regions indicate periods of time when a province was not in existence.}\label{fig:all-cases}
\end{figure}

\end{knitrout}

When forecasting, we will only ever have the cases recorded prior to the time the predictions are made. So that we could compare the expected real-time performance of models as if they had been applied in real-time, all data were archived in the database with a time-stamp on arrival. 
This enabled researchers to ``turn back the clock'', i.e. to query data that was available at a particular point in time.  
%Since all case reports have some delay between symptom onset and their delivery date, case counts would be different for the queries ``all cases with symptom onset prior to 2014-04-01'' and ``all cases reported before (i.e. with delivery date prior to) 2014-04-01''.  %In surveillance systems that may have short times between symptom onset and case report (e.g. Singapore's all-digital surveillance system reportedly has minimal reporting delays for cases \cite{Hii:2012wi}) this concern may be less of an issue. %But as described below, reporting delays in the Thai surveillance system differ by province, and case reports may take months to find their way into the official surveillance record. Therefore, having a system that measures these delays systematically and can base forecasts on data available at a particular time may play a vital role in creating and evaluating actionable forecasts.
Throughout this manuscript, we use the term ``nowcast'' to refer to predictions made for timepoints on or prior to the analysis date and ``forecasts'' to refer to predictions made for timepoints at or after the analysis date. 

\subsection{Accounting for delays in case reporting}

A key property of a surveillance system is the reporting delay, defined for our purposes as the duration of time between symptom onset and the case being available for analysis. 
%In constructing real-time forecasts, characterizing the reporting delays and what data could be expected to be available at each particular time was a vitally important part of our data exploration process. 
%It can be difficult or impossible to glean from static, historical records of disease incidence (such as, e.g., annual reports that contain weekly incidence numbers), what data were available at a particular time.  %The process of reporting case data varies widely depending on the surveillance system in question. 
%For systems with networked digital record keeping, reporting delays may be minimal, and largely be due to the time it takes individuals to present with symptoms at a doctor's office, clinic, or hospital. In other surveillance systems, the journey of data from local clinic to national surveillance system may take weeks or months. In the U.S., influenza cases aggregated in surveillance data managed by the Centers for Disease Control and Prevention (CDC) often have between [[two and four week]] reporting delays \cite{TK}. 
During 2014 reporting delays for dengue ranged from 1 to 50 weeks. 
This was due to the process of reporting cases. Case reports typically follow a path of reporting from hospitals to district surveillance centers and then to provincial health offices before arriving at the national surveillance center. In all provinces, 50\% of cases were reported within 
6 
weeks and 75\% of cases were reported within 
10 
weeks. However, a small fraction of cases took quite a bit longer.
To account for reporting delays, our models specified a reporting lag $l$, in biweeks. Data with onset dates within last $l$ biweeks was considered underreported and left out from the analysis. %For the predictions presented here, we fitted models to data with lags of 6 biweeks. 
We present results from the models with a lag of 6 biweeks (about 3 months), as these produced stable predictions across the entire country. %and showed in subsequent analyses to have comparable aggregate performance with full-data forecasts that started at the same timepoint (see Table \ref{tab:absolute-horizon}).
More sophisticated adjustments for reporting delays are the subject of our team's ongoing research.
%To leverage all of the available data at a given time, methods could be developed to impute the remaining number of yet-to-be-reported cases in these recent biweeks using past observations on reporting delays. Correcting for reporting delays is the subject of our team's ongoing research.

\subsection{Timing of predictions}
While the predictions presented in this manuscript were made retrospectively, in 2015 when complete data were available, they were constructed to mimic real-time predictions by using only the data available at each analysis date in 2014. During the 2014 calendar year, predictions from a similar model were generated in real-time and disseminated to the Thai Ministry of Public Health. We chose the set of analysis dates as the first day of each biweek for which data had been delivered in the previous biweek (Supplementary Table 2). 
For each analysis date in 2014, we used the candidate model to generate ``real-time'' province-level biweekly predictions for the subsequent 10 biweeks (5 months).

\section{Methodology for predicting case counts}

\subsection{Disease model: features and estimation}

\subsubsection*{Statistical model}
We assumed the biweekly province-level reported cases follow a Poisson distribution, where the previous biweek's reported cases serve as an offset term. 
Let the number of cases with onset occurring within time interval $t$ in province $i$ be represented as a random variable $Y_{t,i}$, then
\begin{eqnarray*}
Y_{t,i} & \sim & Poisson(\lambda_{t,i} \cdot y_{t-1,i})
\end{eqnarray*}
where the lag-1 term $y_{t-1,i}$ is used as an offset in this model. We adopt the convention of using lower-case $y_{t,i}$ to indicate previously observed case counts that are treated as fixed inputs in our model.  We explicitly model the rate $\lambda$ as 
\begin{eqnarray}
\log \lambda_{t,i} & = & f(b(t)) + g(t) + \sum_{j\in \mathcal{C}} \sum_{k \in \mathcal{L}} \alpha_{j,k} \log \frac{y_{t-k,j}+1}{y_{t-k-1,j}+1} \label{eq:spamd-model}
\end{eqnarray}
where $\mathcal{C}$ is the set of $J$ most-correlated provinces with province $i$ and $\mathcal{L}$ is the set of lag times used in the model; $b(t)$ is the biweek of time $t$; $f(b(t))$ is assumed to be a cyclical cubic spline with period of one year (i.e. 26 biweeks); and $g(t)$ is a smooth spline to capture secular trends over time. Adding 1 to the numerator and denominator of the correlated province covariates ensures that the quantities are defined when no case counts are observed at a particular province-biweek. This method of adjusting for zero counts has been interpreted as an ``immigration rate'' added to each observation \cite{Zeger:1988up}. 

We note that the model can be expressed as 
\begin{equation}
\lambda_{t,i} = \mathbb{E}\left [ \frac{Y_{t,i}}{y_{t-1,i}} | y_{t-1,i} \right ] \approx R_{t,i} \label{eq:spamd-Rt}
\end{equation}
which shows that $\lambda_{t,i}$ can be interpreted as the expected reproductive rate at time $t$ in location $i$, or $R_{t,i}$ \cite{Nishiura:2010jh}.

These models were fit using the Generalized Additive Model (GAM) framework (i.e. as generalized linear models with smooth splines estimated by penalized maximum likelihood) \cite{Hastie:1990vg}, using the mgcv package for R \cite{wood2011, ihaka1996r}. Each province's time-series was subset to remove any cases from the previous $l$ biweeks. The remaining data were smoothed before fitting the model and making predictions.

Seasonal patterns were modeled using a penalized cubic regression spline, constrained to have a cycle of one year with continuous second derivatives at the endpoints.
Secular trends were modeled using penalized cubic splines with 5 equally spaced knots over 47 years (roughly one knot per decade).

%{\bf Population dynamics}\\
Information on epidemic progression elsewhere in the country was taken into account by including reported case counts at 1 lagged timepoint for the 3 most correlated provinces with province $i$ in the data used to fit the model.  Details of this model selection are provided in the appendix.

We approximated the predictive distribution for all provinces using sequential stochastic simulations of the joint distribution of the case counts for each province. Prediction intervals were generated by taking quantiles (e.g., the 2.5\% and 97.5\%) of this distribution. Full details of the methods used to generate the multi-step predictions are available in the appendix. 
%For example, if data through time $t^*$ is used to fit the models for all locations, then a single simulated prediction consists of a simulated Markov chain of dependent observations for time points $t^*+1$, $t^*+2$, ..., $t^*+h$, across all provinces. Due to the interrelations between the provinces, as shown in equation \ref{eq:spamd-model}, we simulate counts for all provinces at a single timepoint before moving on to the next timepoint. We created $M$ independently evolving sequential predictions by, at each prediction time point, drawing from each province-specific Poisson distribution with means given by equation \ref{eq:spamd-model}. For a given prediction horizon $h$, this process generates an empirical posterior predictive distribution. With $M$ sufficiently large, this distribution can be used to construct $p$-\% prediction intervals for each province-timepoint by taking the $\frac{100-p}{2}$ and $1-\frac{100-p}{2}$ quantiles of the empirically simulated distribution.

\subsection{Metrics for evaluating predictions}
We used several different metrics for evaluating our predicted case counts. We calculated Spearman correlation coefficients to measure the agreement between predicted and observed values. We also calculated the mean absolute error (MAE) by aggregating across analysis times within a given province. We computed the relative mean absolute error (relative MAE) comparing the predictions for a given province to predictions from a seasonal average baseline model. The seasonal baseline model for a given province is the median value of previously observed counts for the given biweek in that province over the past 10 years. The use of absolute error metrics over squared error metrics has been encouraged to enhance interpretability \cite{Hyndman:2006dt,Reich:2AH3z5zJ}. Additionally, we calculated empirical 95\% prediction interval coverage as the fraction of times the 95\% prediction interval covered the true value.
% \begin{itemize}
% \item Correlation coefficients
% \item Relative MAE vs. seasonal and vs. last observation
% \item Total observed annual cases vs. predictions from pre-high season
% \item CI coverage
% % \item Number of biweeks with an outbreak during high season
% % \item For provinces with outbreaks, the sensitivity of outbreak predictions 0, 2, and 4 weeks ahead of time
% % \item For all provinces, the log score for peak biweek projections, by analysis date
% % \item Whether or not the model correctly predicted the peak week (and +/- k weeks)
% \end{itemize}

\subsection{Real-time vs. full-data predictions}
We evaluated the performance of our real-time forecasts against predictions that could have been made had a full dataset been available at the analysis dates. To make this comparison, we ran a set of multi-step forecasts for 2014 at each analysis date using the complete data for 2014 that was finalized in late April 2015.
This experiment was designed to focus on two comparisons. First, we aimed to compare real-time and full-data predictions where the multi-step predictions began at the same timepoint. %This analysis addressed the question of how much the real-time predictions were impacted by the underreported data once we accounted for the reporting delays by removing the most recent 6 biweeks of data. 
Second, we aimed to compare, by horizon, the real-time and full-data predictions where the origin of the multi-step full-data predictions was anchored at the analysis time but the origin of the real-time predictions was 6 biweeks earlier to account for underreported data. %This analysis addressed the question of how much better or worse our model would have done if we did not need to adjust for underreported data by dropping the past 6 biweeks.

\begin{knitrout}
\definecolor{shadecolor}{rgb}{0.969, 0.969, 0.969}\color{fgcolor}\begin{figure}
\includegraphics[width=\maxwidth]{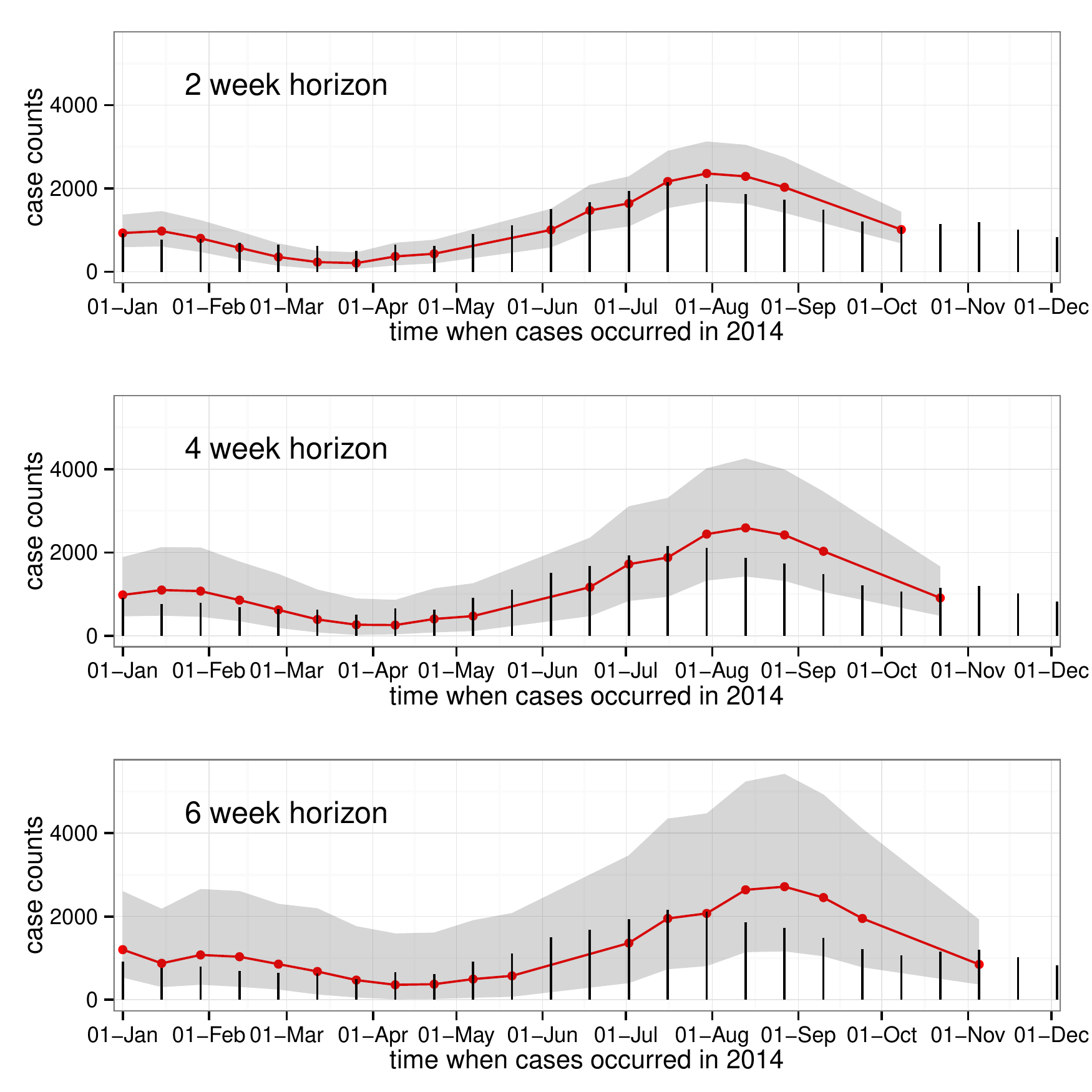} \caption[Country-wide real-time predictions for incident dengue hemorrhagic fever]{Country-wide real-time predictions for incident dengue hemorrhagic fever. Red lines show predicted case counts, black bars show cases reported by the end of the 2014 reporting period. The three figures show (top to bottom) one-, two-, and three-biweek ahead predictions. So, for example, every dot on the top graph is a one-biweek ahead real-time prediction made from all available data at the time of analysis.}\label{fig:step1-forecasts}
\end{figure}

\end{knitrout}

\begin{knitrout}
\definecolor{shadecolor}{rgb}{0.969, 0.969, 0.969}\color{fgcolor}\begin{figure}
\includegraphics[width=\maxwidth]{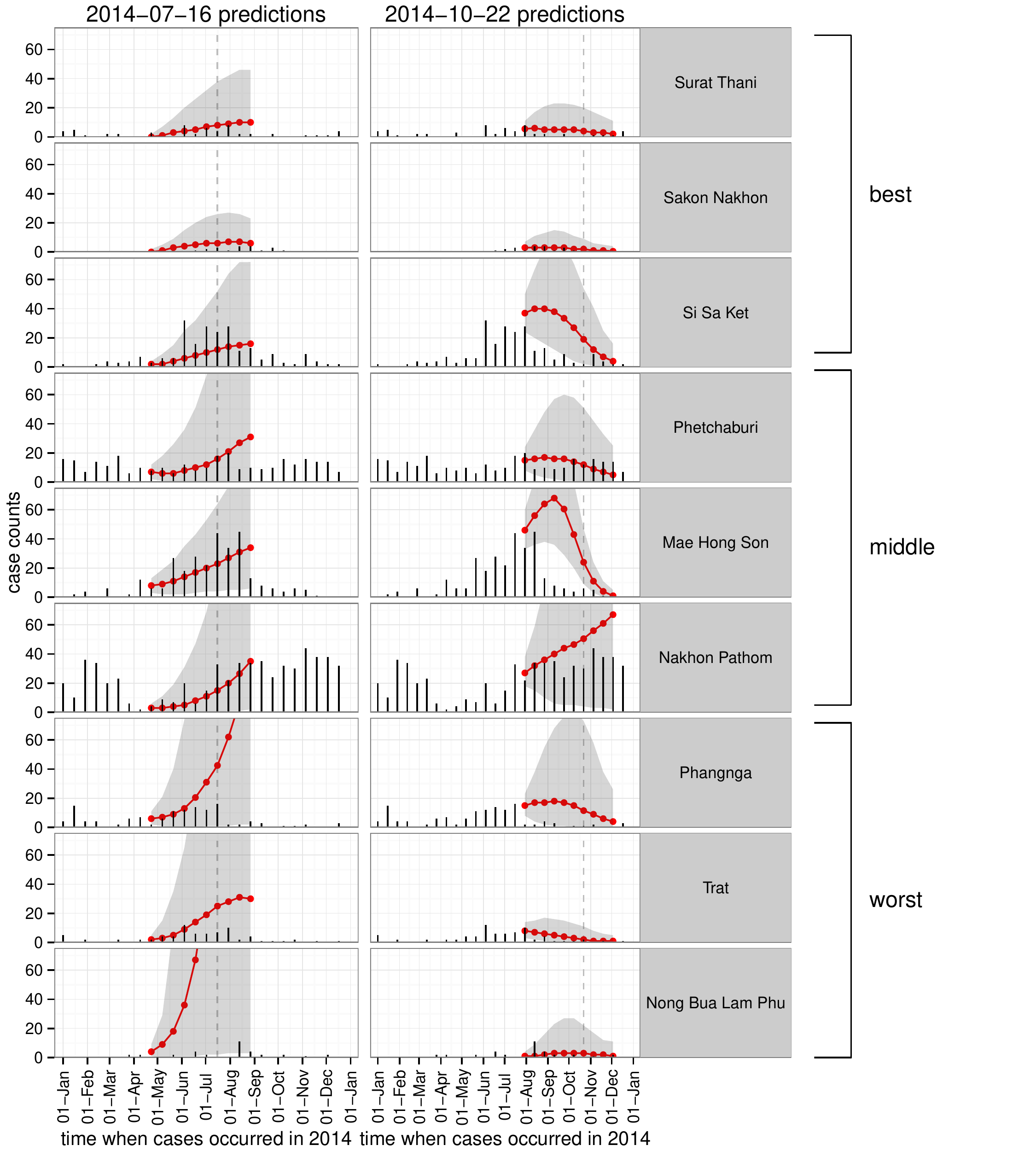} \caption[Ten-step forward predictions made at two time-points in 2014]{Ten-step forward predictions made at two time-points in 2014. Results for nine provinces are shown, representing (from top to bottom) the best three provinces, the middle three, and the worst three performing provinces in terms of relative mean absolute error when compared to a seasonal reference model.}\label{fig:province-specific-forecasts}
\end{figure}

\end{knitrout}

\begin{knitrout}
\definecolor{shadecolor}{rgb}{0.969, 0.969, 0.969}\color{fgcolor}\begin{figure}[H]
\includegraphics[width=\maxwidth]{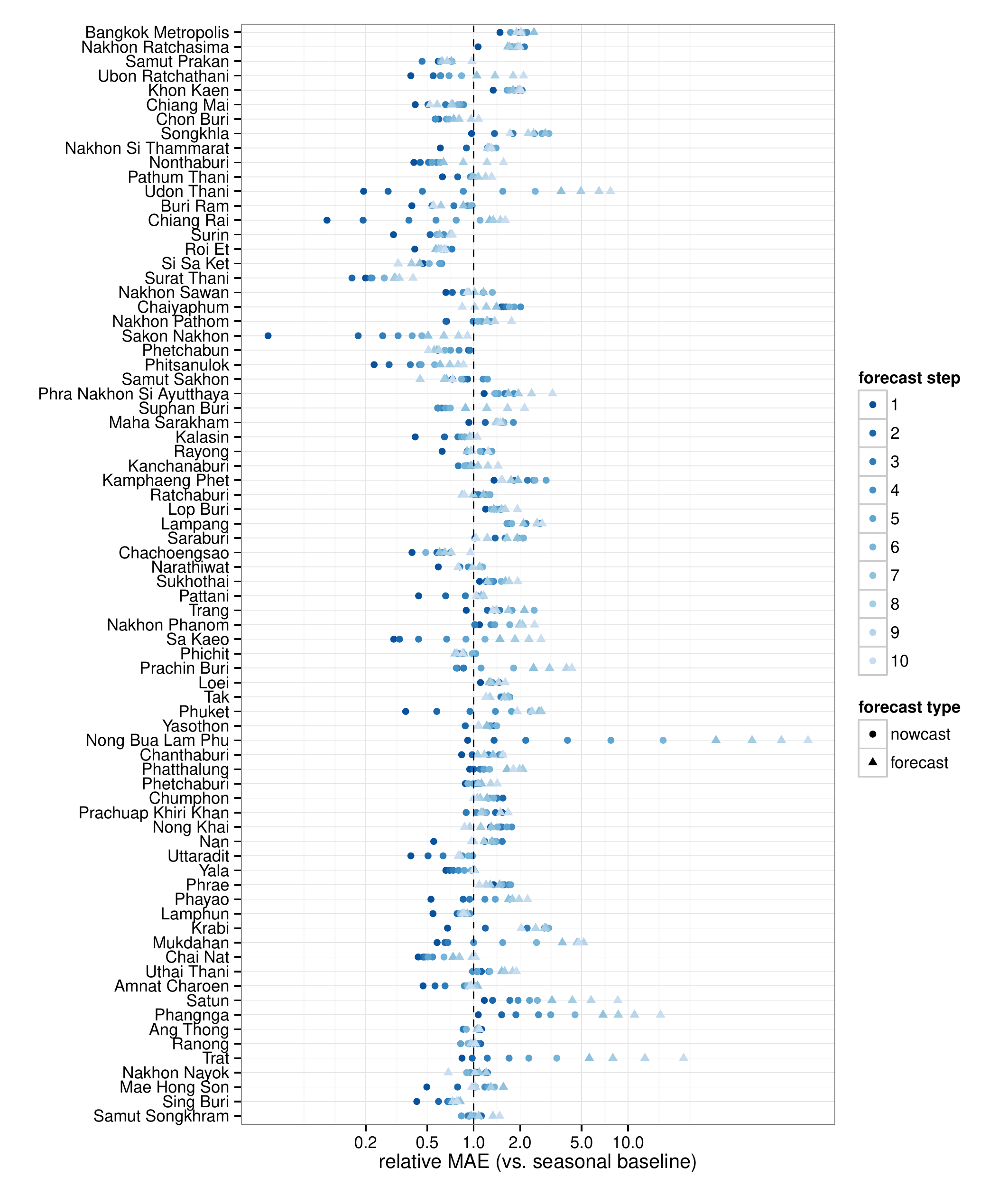} \caption[Relative mean absolute error (MAE) comparing our prediction model vs]{Relative mean absolute error (MAE) comparing our prediction model vs. a model that predicts a seasonal average, by province and step forward (in biweeks).}\label{fig:relative-mae}
\end{figure}

\end{knitrout}

% latex table generated in R 3.2.1 by xtable 1.7-4 package
% Sun Nov 15 22:21:21 2015
\begin{table}[ht]
\centering
\begin{tabular}{c|cc|ccccc}
   &&& \multicolumn{5}{|c}{relative MAE (real-time vs. seasonal baseline)} \\horizon (h) & $R^2$ & 95\% PI coverage & $Q_5$ & $Q_{25}$ & $Q_{50}$ (median) & $Q_{75}$ & $Q_{95}$ \\ 
  \hline
1 & 0.92 & 0.61 & 0.22 & 0.46 & 0.70 & 1.07 & 1.49 \\ 
  2 & 0.90 & 0.93 & 0.28 & 0.60 & 0.90 & 1.25 & 1.79 \\ 
  3 & 0.89 & 0.99 & 0.43 & 0.68 & 0.96 & 1.48 & 2.18 \\ 
  4 & 0.84 & 0.99 & 0.56 & 0.83 & 1.09 & 1.42 & 2.45 \\ 
  5 & 0.79 & 0.99 & 0.53 & 0.83 & 1.16 & 1.55 & 2.81 \\ 
  6 & 0.73 & 1.00 & 0.55 & 0.87 & 1.22 & 1.70 & 3.02 \\ 
  7 & 0.65 & 0.99 & 0.58 & 0.88 & 1.22 & 1.67 & 3.70 \\ 
  8 & 0.57 & 0.99 & 0.61 & 0.94 & 1.16 & 1.72 & 4.73 \\ 
  9 & 0.52 & 0.99 & 0.57 & 0.95 & 1.21 & 1.83 & 5.93 \\ 
  10 & 0.50 & 0.99 & 0.54 & 0.91 & 1.27 & 1.92 & 7.91 \\ 
  \end{tabular}
\caption{Summary of real-time prediction accuracy, by prediction horizon. These results are aggregated across all provinces. The $R^2$ and 95\% PI coverage columns present the overall correlation coefficient and prediction interval coverage. The relative MAE columns show five quantiles of the distribution of province-level relative MAEs comparing the real-time model at the given horizon to a seasonal baseline model at the given horizon: $Q_5$ (the 5$^{th}$ percentile), $Q_{25}$ ($25^{th}$ percentile), $Q_{50}$ (median), $Q_{75}$ ($75^{th}$ percentile), and $Q_{95}$ (the 95th percentile). The relative MAEs were calculated as the MAE from the real-time predictions divided by the MAE from the seasonal average predictions, therefore values larger than 1 indicate that the real-time models showed more absolute error on average than the seasonal models.} 
\label{tab:results-by-horizon}
\end{table}

\section{Forecast results from 2014}

\subsection{Summary of province-level forecasts}

%[Absolute error summary] 
In general, the model predictions showed good, if overconfident, performance at short horizons but less accuracy and high uncertainty at longer horizons. Across all provinces, the correlation between observed and predicted values was 0.92 at a horizon of 1 biweek (2 weeks) and 0.5 at a horizon of 10 biweeks, or approximately 5 months (see Table \ref{tab:results-by-horizon}). Across all provinces, observed 95\% prediction interval coverage was lower than expected at horizons of 2 and 4 weeks (61\% and 93\%, respectively), showing that the models were overconfident in their short-term predictions. This prediction interval coverage increased to 99\% at a 6 week (3 biweek) prediction horizon, and remained at that level for longer horizons. This indicates that our models often had an abundance of uncertainty at mid- and long-term horizons. Figure \ref{fig:step1-forecasts} shows case counts and predictions aggregated across all provinces at horizons of 2, 4, and 6 weeks (or 1, 2, and 3 biweeks).   

Figure \ref{fig:province-specific-forecasts} shows examples of multi-step predictions from two analysis dates in 2014. We show the results from nine distinct provinces, representing the best three provinces, the middle three provinces, and the worst three provinces in terms of relative MAE when compared to a seasonal reference model. The increasing uncertainty is visible in many cases, even when the point-predictions remain close to the true values. The explosive forecasts tended to occur more frequently in the early- and mid-season, when the historical seasonal trend rises and when the observed case counts tend to be increasing from one biweek to the next.

%[Relative error summary]  
There was substantial variation in predictive performance across provinces. Figure \ref{fig:relative-mae} shows the relative mean absolute error (relative MAE) of model predictions compared to a seasonal baseline model at prediction horizons of 2 through 20 weeks (1 through 10 biweeks). We note that predictions during the first three months are nowcasts, as the most recent 6 biweeks of data are ignored in the fitting process and predictions are made starting from the point at which full data was assumed. 

To compare predictive performance of our model between provinces, we used the relative MAE with a simple seasonal model as a baseline. Table \ref{tab:results-by-horizon} summarizes relative MAEs by prediction horizon.  Relative to seasonal baseline prediction models, a majority of provinces made better predictions on average than the seasonal model at a 2, 4, and 6 weeks (i.e. 1, 2, and 3 biweeks) prediction horizons (i.e. up to 1.5 months from the starting point of the predictions). Up to about 5 months from the origin of the multi-step predictions (and two months from the analysis time), over 25\% of province-specific models made predictions that were on average better than the seasonal baseline model. Some province-specific models showed substantially worse predictions when compared to a seasonal baseline at these longer prediction horizons (Table 2). No single province feature (e.g. total average cases, strength of seasonal trends, population size, season-to-season variation) was able to explain the substantial variations in performance, highlighting the challenges of creating a unified modeling framework for a set of varied locations (see appendix for details).   

%[Model diagnostic summary] 
% Despite treating the past 6 biweeks (approximately 3 months) as completely unobserved data, many provinces showed sensitivity to partially observed case counts serving as covariates for prediction.  Explosive forecasts were observed  ... [Give some stats on when and where the explosive forecasts happened]

% latex table generated in R 3.2.1 by xtable 1.7-4 package
% Sun Nov 15 22:21:22 2015
\begin{table}[ht]
\centering
\begin{tabular}{c|ccccc}
   & \multicolumn{5}{|c}{relative MAE (real-time vs. full-data baseline)} \\horizon & $Q_5$ & $Q_{25}$ & $Q_{50}$ (median) & $Q_{75}$ & $Q_{95}$ \\ 
  \hline
1 & 0.82 & 0.92 & 0.98 & 1.03 & 1.13 \\ 
  2 & 0.79 & 0.91 & 0.98 & 1.01 & 1.16 \\ 
  3 & 0.77 & 0.91 & 0.96 & 1.01 & 1.17 \\ 
  4 & 0.77 & 0.91 & 0.97 & 1.01 & 1.10 \\ 
  5 & 0.83 & 0.90 & 0.97 & 1.00 & 1.10 \\ 
  6 & 0.82 & 0.93 & 0.99 & 1.03 & 1.11 \\ 
  7 & 0.82 & 0.94 & 1.00 & 1.06 & 1.16 \\ 
  8 & 0.82 & 0.95 & 1.03 & 1.09 & 1.20 \\ 
  9 & 0.84 & 0.93 & 1.03 & 1.13 & 1.28 \\ 
  10 & 0.85 & 0.96 & 1.05 & 1.18 & 1.36 \\ 
  \end{tabular}
\caption{Comparison of province-level prediction accuracy between full-data and real-time predictions, by prediction horizon. These calculations assume that both the full-data and real-time multi-step predictions began at the same time. The table shows the 5th percentile ($Q_5$), 25th percentile ($Q_{25}$), median ($Q_{50}$), 75th percentile ($Q_{75}$), and 95th percentile  ($Q_{95}$) value of the relative MAE from each province at the given horizon. The relative MAEs were calculated as the MAE from the real-time predictions divided by the MAE from the full-data predictions, i.e. values larger than 1 indicate that the real-time models showed more absolute error on average than the full-data models.} 
\label{tab:fulldata-horizon}
\end{table}

% latex table generated in R 3.2.1 by xtable 1.7-4 package
% Sun Nov 15 22:21:22 2015
\begin{table}[ht]
\centering
\begin{tabular}{c|ccccc}
   & \multicolumn{5}{|c}{relative MAE (real-time vs. full-data baseline)} \\absolute horizon & $Q_5$ & $Q_{25}$ & $Q_{50}$ (median) & $Q_{75}$ & $Q_{95}$ \\ 
  \hline
1 & 0.92 & 1.21 & 1.49 & 2.05 & 6.62 \\ 
  2 & 0.76 & 0.98 & 1.27 & 1.79 & 6.40 \\ 
  3 & 0.66 & 0.91 & 1.19 & 1.88 & 5.45 \\ 
  4 & 0.55 & 0.90 & 1.32 & 1.87 & 4.96 \\ 
  \end{tabular}
\caption{Comparison of province-level prediction accuracy between full-data and real-time predictions, by prediction horizon. These results were computed comparing predictions as if the full data was available at the analysis time with the real-time predictions that build in a 6-biweek (approximately 3 month) buffer to account for delayed case data. The table shows the 5th percentile ($Q_5$), 25th percentile ($Q_{25}$), median ($Q_{50}$), 75th percentile ($Q_{75}$), and 95th percentile  ($Q_{95}$) value of the relative MAE from each province at the given horizon. The relative MAEs were calculated as the MAE from the real-time predictions divided by the MAE from the full-data predictions, i.e. values larger than 1 indicate that the real-time models showed more absolute error on average than the full-data models.} 
\label{tab:absolute-horizon}
\end{table}

\subsection{Comparing real-time to full-data predictions}
We compared real-time  and full-data predictions that began at the same timepoint. This analysis can help answer the question of how much the real-time predictions were impacted by the underreported data, once accounting for the reporting delays by removing the most recent 3 months of data.
As shown in Table \ref{tab:fulldata-horizon}, these analyses demonstrated that once went back 3 months to begin the nowcasting, more than 50\% of the provinces had more accurate real-time forecasts than full-data forecasts at all prediction horizons up to 3 months in advance. This suggests that inaccuracies in the real-time predictions once those recent 3 months are discarded are driven less by the reporting delays than they are by model misspecification and other background noise in the data.
%Additionally, we observed no clear relationship between the duration of reporting delays in a province (measured by the fraction of cases delivered after 6 biweeks) and the relative MAE between real-time and full-data predictions.

A second analysis compared real-time predictions with a horizon of 7 biweeks with full-data predictions at 1 biweek. This analysis can tell us how much better or worse our model would have done if we did not need to adjust for underreported data by dropping the past 3 months, i.e. if all of our data were available at the time of analysis. We refer to this realignment of horizons as the absolute horizon, to suggest that a real-time prediction that removes 6 biweeks of data and then projects 7 steps forward is predicting the same timestep as a full-data prediction that does not remove any data and just projects 1 biweek forward. Results from this analysis are shown in Table \ref{tab:absolute-horizon} for absolute horizons of 1 through 4 biweeks. Overall, 66 of the 76 provinces (87\%) showed better average performance in the full-data forecasts at 1 step ahead than the real-time forecasts at 7 steps ahead (i.e. had a relative MAE of greater than 1). In a majority of the provinces at each absolute horizon the full-data forecasts were on average closer to the true value than the real-time forecasts. However across all the absolute horizons, for between 10 and 26 provinces the full-data predictions had more error than the real-time predictions. 
While it is surprising that full-data predictions under performed real-time predictions in such a high percentage of the provinces, this reflects the challenges of making predictions in such a noisy system.
%While it was unexpected that multi-step ahead predictions from under-reported data could be consistently more accurate than a single step ahead prediction from fully observed data, these results again reflect the difficulty of even making single-step-ahead predictions in noisy systems. 
A sample of predictions by province and analysis date are provided as supplemental figures to illustrate this challenge.

\section{Discussion}
We present the prediction results from our real-time prediction infrastructure established for dengue hemorrhagic fever in Thailand. This infrastructure addresses several key operational features of real-time predictions, including real-time data management, the impact of reporting delays, and incorporating a disease transmission model that takes into account spatial and temporal trends. 

The infectious disease prediction literature has a rich and varied selection of prediction algorithms but has not historically focused on the challenges of operationalizing predictions in real-time. Continued development and refinement of such prediction pipelines, such as that presented here, will enable existing prediction methods to reach their full potential in making an impact on public health decision-making and planning.

The infrastructure that we have developed for integrating real-time data into predictions for the Thai Ministry of Public Health is the result of a long-standing governmental/academic partnership between the Ministry and U.S.-based researchers. This collaboration has enabled the creation of a single, unified authoritative source of almost all governmental dengue surveillance ever collected in Thailand, dating back nearly 50 years \cite{Limkittikul:2014dh}. Additionally, by enabling the transmitting of surveillance data in near real-time (every two weeks from October 2013 and continuing through the writing of this manuscript 2015), this effort has created a valuable dataset that has catalogued the reporting delays in a live surveillance system. %Our team is working on developing additional methodology to more actively use this data to impute the number of unreported cases. %These data could be used in future efforts to understand and characterize the workings of this long-standing and well-respected disease surveillance system.\cite{Limkittikul:2014dh} The shared vision of our collaborative team is that predictions generated from this data-to-predictions pipeline will be used to inform public health decision making in Thailand beginning with the 2016 calendar year. 

Formal data archiving protocols should be followed when making real-time predictions. Real-time predictions should be (1) generated prior to having the final data available and (2) formally registered or time-stamped in an independent data repository. Taking these steps ensures that no bias (intentional or not) enters the scientific process of evaluating the predictions. 

While we are actively developing and validating other prediction models for this data, we chose to report the results from the prediction model that we used during 2014 to provide draft predictions to Thai public health officials. We intentionally did not perform extensive {\em post hoc} model validation or evaluation to minimize the risk of overfitting our model to this particular dataset.  

Our 2014 real-time predictions varied substantially by province in quality and public health utility. In close to half of the Thai provices, our model out-performed a seasonal baseline model predicting one month in advance. As the horizon moves forward, the seasonal baseline model makes better predictions in more provinces: at a 5 month horizon, just over 25\% of provinces are predicted better by our model than the seasonal model. 

Our ability to make effective predictions into the future in a majority of provinces is made difficult by delayed case reporting. Our analyses show that if there were no reporting delays, our model would make substantially more accurate predictions in over 50\% of the Thai provinces (Table 3).

%Our 2014 real-time predictions varied substantially by province in quality and public health utility. In close to half of the Thai provinces, our model outperformed a seasonal baseline model predicting one month in advance. As the horizon moves forward, the seasonal baseline model makes better predictions in more provinces: at a 5 month horizon, just over 25\% of provinces are predicted better by our model than the seasonal model. Our ability to make effective predictions into the future in a majority of provinces is made difficult by delayed case reporting. Our analyses show that if there were no reporting delays, our model would make substantially more accurate predictions in over 50\% of the Thai provinces (Table 3).
% calculated as 100*(1-1/6.6) and 100*(1-1/1.2)

%In our current framework, this means that we must begin making predictions by starting three months in the past, using a single model to create now-casts (predictions that bring us up to the current time) and then forecasts (making predictions about observations at future timepoints). In our current framework, the nowcasting and forecasting models are one and the same, although more generally these models need not be the same.

While we have conducted extensive evaluation of the performance of our real-time predictions in 2014, this may not represent the performance of the model in other years. There is substantial year-to-year variation in annual province-level incidence in Thailand. The annual total number of cases observed in 2014 were in the lower half of previously observed annual incidence in 49 of 76 provinces. A complete characterization of our real-time model's predictive performance will require evaluation across multiple years of data that is arriving in real-time, or with historical complete data with synthetically created reporting delays. %Since we only know the reporting delays for cases delivered in the last two years, we are limited in our ability to simply assess the model performance with data from previous years. However, this process could be accelerated via simulation studies that use observed reporting delays to create synthetic underreported data for previous year's data.

The simplicity of the statistical prediction models that we present in this manuscript are both a strength and a weakness. This type of phenomenological time-series model tends to show good predictive performance in the short term but have known deficiencies when making long-term predictions. Additionally, when forecasting forward from auto-regressive models, this can lead to instabilities and explosive forecasts, as was observed in the predictions from some of the provinces. Also contributing to the instability of our models in a prediction context are that we do not incorporate uncertainty in and use a smoothed value of the $y_{i, t-1}$ offset term. %Furthermore, while our results from the full-data and real-time prediction comparison suggest that on average the predictions are not made worse by using only the partially observed data, we are not currently modeling the underreported counts, instead relying on the assumption that cases are fully reported after three months (6 biweeks). We know to be an optimistic assumption in many settings and this may contribute to instability and noise in the model covariates and offset.

The model that we present here has been shown to perform well in contexts where there are no reporting delays [manuscript in preparation]. The auto-regressive model used in this work is based on a standard statistical auto-regressive integrated moving average (ARIMA) models. In fact, the reformulation of the ARIMA model in a disease transmission model context -- making explicit the connection between modeling auto-regressive counts and the reproductive number, as shown in equation \ref{eq:spamd-Rt} -- is an important link between commonly used models in different fields. %Possible model improvements could include regularizing seasonality by distance between provinces, choosing the correlated provinces serially through partial correlations, and incorporating overdispersion of case counts.

%[The state of and challenges and frameworks for multi-step predictions (cite Ben Taieb's work).]
%Adding additional sophisticated statistical learning machinery (e.g. ensemble prediction models) could improve the performance of our predictions. Ensemble models are frequently among the top competitors in prediction competitions, and have a strong reputation in the machine learning and statistical communities for generating robust classifications and predictions. However, they have been less frequently used in regression-style analyses, and are virtually unproven for multi-step forecasts of highly correlated time-series data – the type often observed in infectious disease situations.\cite{BenTaieb:2014vg} Machine learning research has suggested that ensemble models that create direct forecasts of all time-steps may have better predictive performance than recursive methods \cite{BenTaieb:2012in,Taieb:2015cu}. Our current model uses a recursive method for generating preditions. %could be used as one of several inputs into an ensemble model.

%[The state of real-time infectious disease forecasting]
%Registering real-time predictions to protect against reporting/publication bias towards better predictions.

The past decade of biomedical research has borne witness to rapid growth in digital surveillance data. A pressing challenge for the professional and academic epidemiological and biostatistical communities is to learn how to turn this deluge of data into evidence that informs decision making about improving health and preventing illness at the individual and population levels. Improved real-time forecasts of infectious disease outbreaks can inform targeted intervention and prevention strategies. Continued research and collaboration in this area will lead to a better understanding of how to integrate infectious disease predictions into public health practice. 
The collaborative effort described by this manuscript provides a template for operationalizing real-time predictions and describes specific results from this effort to integrate modern tools of data science with public health decision making.

%[The vision for incorporating these predictions into real-time public health decision making]

%Add connections to ARMA model in using rates.

%    \item "Dengue disease outbreak definitions are implicitly variable" \cite{Brady:2015jl}

% Possible improvements:
% \begin{itemize}
%     \item more standard arima model?
%     \item regularize seasonality by distance between provinces
%     \item choose correlations serially
%     \item overdispersion of counts
% \end{itemize}

\section*{Acknowledgments}
\begin{enumerate}
\item This work was supported by the National Institute of Allergy and Infectious Diseases at the National Institutes of Health (grants R21AI115173 to NGR and DATC and R01AI102939). 
\item Department of Biostatistics and Epidemiology, School of Public Health and Health Sciences, University of Massachusetts - Amherst, Amherst, Massachusetts (Nicholas G Reich, Stephen A Lauer, Krzysztof Sakrejda); 
Office of Disease Prevention and Control 1, Bangkok, Thailand (Sopon Iamsirithaworn); 
Bureau of Epidemiology, Department of Disease Control, Ministry of Public Health, Bangkok, Thailand (Soawapak Hinjoy, Paphanij Suangtho, Suthanun Suthachana);
Department of Epidemiology, Johns Hopkins Bloomberg School of Public Health, Baltimore, Maryland (Hannah E Clapham, Henrik Salje, Derek A T Cummings, Justin Lessler); and, 
Emerging Pathogens Institute, Department of Biology, University of Florida, Gainesville, Florida (Derek A T Cummings).
\end{enumerate}

\bibliographystyle{unsrt}
\bibliography{dengue-real-time}

\end{document}